\documentstyle{article}
\newcommand{\be}{\vskip 0em plus 0.1em minus 0.5em\begin{eqnarray}}
\newcommand{\ee}{\end{eqnarray}\vskip 0em plus 0.1em minus 0.5em}
\newcommand{\nn}{\nonumber \\}
\begin{document}
\title{Positronium as an example of algebraic\\
       composite calculations}
%\subtitle{How to build an algebraic approach to composites}
\author{Bertfried Fauser\thanks{
        E--Mail:~Bertfried.Fauser@uni-tuebingen.de}
        and Harald Stumpf\\
        Institut f\"ur Theoretische Physik\\
        Auf der Morgenstelle 14\\
        D--72076 T\"ubingen}
%
%\end{opening}
%
%\begin{center}Submitted, October 25\end{center}
\date{Submitted, October 25\\
 Tue-prep-95-10-08; hep-th/jjmmnnn}
\maketitle
\hyphenation{}
\begin{abstract}
The functional quantum field theory, developed by Stumpf,
provides the possibility to derive the quantum dynamics of a
positronium gas from Coulomb interacting electrons and
positrons. By this example, the method will be brought in a
Clifford algebraic light, through identifying the functional
space with an infinite dimensional Euclidean Clifford algebra.
\end{abstract}

%
%{
%\def\@makefnmark\relax
%\def\@thefnmark\relax
%\footnote{J. Keller and Z. Oziewicz (eds.), The Theory of The Electron,
%your pages.}
%\footnote{\copyright 1996 Universidad Nacional Autonoma de
%M\'exico. Printed in M\'exico.}
%}
%
\vskip 0em plus 1em minus 1em
\tableofcontents

\section{Weak Mapping Formalism of QED}

\subsection{Motivation}

If one likes to describe physics by atomistic theories, for
instance the standard model in elementary particle physics,
then one has to provide a theory of composites. This is necessary
in order to theoretically build up mesons, molecules etc., i.e.
the atomistic hierarchy. As the standard model is a
parameterization of nature with a large number of parameters and
``elementary'' fields, it is obvious that such a theory needs a
further explanation from first principles, which again leads to
composites.

Current quantum field theoretic methods are based on
perturbation theory. The most outstanding success of such
perturbative methods are the $g-2$--factor and the Lamb shift.
However calculations where composites are involved can {\it not}
be done in a perturbative manner, because of the strong coupling
of the constituents. So other methods have been developed. For
example the operator product method \cite{Operator-pro}. This
method considers a pair of operators as a composite. This
contradicts the field  theoretic point of view, that such a
construct should be composed by an infinity of sub--particles
generated from the vacuum polarization cloud. In addition,
quantum numbers of states can not be uniquely assigned to
operator products beyond perturbation theory. Another method to
describe composites are Bethe--Salpeter like calculations
\cite{BetSal}. These calculations are based on perturbation
theory and can not be used to calculate for example exited
states for the hydrogen atom \cite{BSFlaw}. Even if one would
succeed to calculate the correct  energy levels, one has not
achieved a dynamical theory of the bound states.

In this note, we want to outline the quantum field theoretic
method of "weak mapping", which was developed in order to
non perturbatively treat the composite particle effective
dynamics. We use as an example the Quantum--Electro--Dynamics
(QED). This method is formulated in terms of quantum field
theoretic quantities and makes use of the technique of
generating functionals \cite{StuBor}. However, besides the
necessity of giving a short outline of the method, we are mostly
interested in the connection of quantum field theories to
Clifford algebras. The link between these  quite different
schemes are the generating functionals. The quantum field
theoretical approach involves operators, commutators,
propagators, states etc. and tries to give a consistent scheme
based on them. The Clifford algebra consists from the very
beginning of geometrical entities and geometrical relations,
which provide a built in interpretation in geometrical terms. The
task of a transition between these pictures is to translate the
quantum field theoretic considerations into purely geometric
ones and vice versa. It is the hope, that it will become
possible to give an algebraic definition of the composite,
rather than the heuristic ones used up to now. Therefore it is
very useful to have the weak mapping theory available in the
algebraic picture.

\subsection{QED in Transverse Coordinates}

Since QED is a gauge theory, we have to deal with a constrained
theory. Especially on the quantum level this is a difficult
task \cite{QED-neu}. Hence in a first step we will eliminate the
gauge freedom and perform the quantization afterwards. This is
only possible in QED, because of the simple Abelian gauge
coupling. In non abelian Yang--Mills Theories as QCD one has a
nonlinear term in the Gauss' law, and a classical elimination
leads to difficulties.

Thus we start with the equations of classical spinor ED
\index{Spinor electrodynamics, covariant}
\be\label{qed-constr}
i)  &&\partial_\nu F^{\mu\nu}+\frac{ie_0}{2}
      \Psi C\gamma^{\mu}\sigma^2\Psi=0 \nn
ii) && (i\gamma^\mu\partial_\mu-m_0)\Psi
       +e_0 A_\mu\gamma^\mu\sigma^3\Psi=0\nn
iii)&&F^{\mu\nu}:=\partial^\mu A^\nu-\partial^\nu A^\mu,
\ee
where we have used the definition \index{Definition, of
super--index}
\be
\Psi &=& \Psi_{\alpha\Lambda}=\left\{
\begin{array}{ll}
\Psi_\alpha & \Lambda=1 \\
\Psi^c_\alpha=C_{\alpha\beta}\bar{\Psi}_\beta & \Lambda=2
\end{array}\right.
\ee
and omit the matrix indices. By definition the sigma matrices
act on the super index $\Lambda$ distinguishing the
field and its charge conjugated field, which has to be treated
as an independent quantity. The gamma matrices and the
charge conjugation matrix $C$ act on the spinor index $\alpha$
of the $\Psi$ field.

Because we are interested in the dynamics, the Hamilton
formalism is convenient \cite{Dir-ham}. This breaks the
explicit Lorentz invariance of the theory. But as we are not
approximating anything, the theory is still
invariant \cite{rel-inv}. The bound state calculation requires an
explicit representation of the binding force. Hence we use the
Coulomb gauge to eliminate the superfluous coordinates. Thus we
set
\be
\partial_k A^k=0&&k\in\{1,2,3\}.
\ee
This enables us to split the `bosonic' fields $A$ and $E$ into
longitudinal and transverse parts. \index{Transverse coordinates
in ED}
\be
A_k\equiv A^{tr}_k,   && \mbox{because of}\quad\partial_k A^k=0 \nn
E_k=E^{tr}_k+E^{l}_k, && E_k:=-F^{0k}=F_{0k}.
\ee
Furthermore we decompose the field equations into dynamic ones
and the remaining constraint, the Gauss' law. \index{Gauss' law}
Looking at (\ref{qed-constr}-i) yields for
\be\label{Gauss}
\nu=0: && i\partial_0 E_k=-\frac{1}{2}e_0\Psi C\gamma^k\sigma^2\Psi
         +i(\partial_j\partial^k A^j-\partial_j\partial^j A^k)\nn
\nu=k: && \partial^k E_k=-i\frac{e_0}{2}\Psi C\gamma_0 \sigma^2\Psi
\ee
and from the definition of $E^k$
\be
i\partial_0 A^{tr,k}=iE^k-i\partial_k A^0.
\ee
With help of the current conservation the Gauss' law
(\ref{Gauss}-$\nu=k$) can now be used  to yield an expression
for the $A_0$ field
\be
A_0=-i\frac{e_0}{2}\Delta^{-1}\Psi C\gamma_0\sigma^2\Psi.
\ee
The $\Delta^{-1}$ symbol is used for the integral kernel of the
inverse of the Laplacian. It is now possible to formulate the
classical spinor ED in generalized, i.e. in independent,
coordinates. For a further compactifying of the notation we
introduce a super field $B_\eta$ also in the `bosonic' fields,
\be
B_k^\eta&:=& \left\{\begin{array}{cl}
A^{tr}_k & \eta=1\\
E^{tr}_k & \eta=2\end{array}\right. .
\ee
The equations of motion then read compactified \index{Spinor
electrodynamics, Coulomb gauged}
\be\label{dyn}
i\partial_0\Psi_{I_1}&=&D_{I_1I_2}\Psi_{I_2}+W^k_{I_1I_2}B_k\Psi_{I_2}
                 +U_{I_1}^{I_2I_3I_4}\Psi_{I_2}\Psi_{I_3}\Psi_{I_4}\nn
i\partial B_{K_1}&=&L_{K_1K_2}B_{K_2}+J_{K_1}^{I_1I_2}\Psi_{I_1}\Psi_{I_2}
\ee
with the definitions
\be
I                  &:=& \{\alpha,\Lambda,{\bf r}\}\nn
K                  &:=& \{k,\eta,{\bf z}\}\nn
P^{tr}             &:=& 1-\Delta^{-1}\nabla\otimes\nabla\nn
D_{I_1I_2}         &:=&-(i\gamma_0\gamma^k\partial_k-\gamma_0
                       m)_{\alpha_1\alpha_2}\delta_{\Lambda_1\Lambda_2}
                       \delta ({\bf r}_1-{\bf r}_2)\nn
W^K_{I_1I_2}       &:=&e_0(\gamma_0\gamma^k)_{\alpha_1\alpha_2}
                       \delta({\bf r}_1-{\bf r}_2)
                       \delta({\bf r}_1-{\bf z})\delta_{1\eta}
                       \sigma^3_{\Lambda_1\Lambda_2}\nn
U_{I_1}^{I_2I_3I_4}&:=&-\frac{i}{8\pi}e_0^2\left[
                       (C\gamma_0)_{\alpha_2\alpha_3}
                       \delta_{\alpha_1\alpha_4}
                       \sigma^2_{\Lambda_2\Lambda_3}
                       \sigma^2_{\Lambda_1\Lambda_4}
                       \frac{\delta({\bf r}_2-{\bf r}_3)
                       \delta({\bf r}_1-{\bf r}_4)}{
                       \vert{\bf r}_1-{\bf r}_2\vert}\right]_{
                       as\{I_2I_3I_4\}} \nn
L_{K_1K_2}         &:=&i\delta({\bf z}_2-{\bf z}_1)\delta_{k_1k_2}
                       \delta_{\eta_11}\delta_{\eta_22}
                       +i\Delta({\bf z}_1)\delta({\bf z}_1-{\bf z}_2)
                       \delta_{k_1k_2}\delta_{\eta_12}
                       \delta_{\eta_21}\nn
J_K^{I_1I_2}       &:=&-\frac{1}{2}e_0 P^{tr}({\bf z}-{\bf r}_1)
                       \delta({\bf r}_1-{\bf r}_2)(C\gamma^k)_{
                       \alpha_1\alpha_2}\delta_{2\eta}
                       \sigma^2_{\Lambda_1\Lambda_2}.
\ee
Now we want to perform the quantization. We therefore impose
commutation relations and the $\Psi$ and $B$ fields become
operators acting on a suitable state space. These states should
belong to an Hilbert space, if the theory would be renormalized. The
topic of renormalization is beyond the scope of this note, and
was discussed in \cite{StuFauPfi} for bound states in a
perturbative manner. As a postulate we introduce the equal time
commutation relations to be \index{Quantization of spinor ED}
\index{QED}
\be\label{comm-rel}
i)  && \{\Psi_{I_1},\Psi_{I_2}\}^t_+:=A_{I_1I_2}=C\gamma_0\sigma^1
       \delta({\bf r}_1-{\bf r}_2)\nn
ii) && [B_K,\Psi_{I}]^t_-:=0\nn
iii)&& [B_{K_1},B_{K_2}]_-^t=:C_{K_1K_2}.
\ee
Equation (\ref{comm-rel}-i) is nothing but the canonical
commutation relation for $\Psi_\alpha$ and
$\bar{\Psi}_\alpha$, the Dirac adjoint field. In our notation
with charge conjugate spinor operators, the somehow unusual
$A_{I_1I_2}$ occurs. Because of the super field notation we have
all four commutators belonging to the $\Psi_\alpha$ and
$\bar{\Psi}_\alpha$ integrated in this relation.

Equation (\ref{comm-rel}-ii) states, that the boson fields are
looked at as elementary fields. That is, the $B$ is {\it not} a
function of the $\Psi_I$ fields. The $B$ field is especially {\it
not} composed of an even number of $\Psi_I$ fields, because these
$\Psi_I$'s should then be out of a disjunct index set $\{I^\prime$\}.
This would be a contradiction, because the relation should hold
for all indices $I$. Equation (\ref{comm-rel}-ii) is usual
postulated in QED.

Relation (\ref{comm-rel}-iii) specifies only the grading of the
boson commutator. That is, the (boson) grade or boson number is
lowered by two through commutation. The result is a generalized
c--number function, thus a complex valued distribution. The
special form of the function $C_{K_1K_2}$ can be calculated as a
consequence of requiring a consistent quantum field theory
\cite{StuFauPfi}, see below (\ref{cons}).

The set of equations (\ref{dyn}) and (\ref{comm-rel}) are the
defining relations of Coulomb gauged quantized spinor QED. All
previous steps are classical and only for convenience to state a
somehow selfconsistent QFT. If one would quantize first and then
try to eliminate the Gauss' law, one would yield an other QFT!
The functional Hamiltonian (see below) would then have infinite
many terms \cite{Lee-QED}.

\subsection{Generating Functionals}

{}From the Wightman reconstruction theorem \cite{Wig} it follows
that if all vacuum expectation values are known, then the field
theory can be reconstructed. Thus one has to give equations for
all such matrix elements. This leads to the
Dyson--Schwinger--Freese equations, for covariant
timeordered matrix elements, which so far can be only derived by
perturbation theory.

If, however, one wants to have a complete quantum field theory
with state space construction and nonperturbative dynamics, this
implies the necessity of using the Hamilton formalism. Therefore
we use one--time functionals exclusively. In our one--time
formalism we can introduce separate orderings for both kinds of
fields. \index{Transition matrix elements}
\be
\tau^t_{nm}(a)&:=&<0\vert{\cal A}(\Psi^t_{I_1}\ldots\Psi^t_{I_n})
                      {\cal S}(B^t_{K_1}\ldots B^t_{K_m})\vert a>.
\ee
Here $<0 \vert$ is the physical vacuum state. Thus both
$\Psi_\alpha$ and $\Psi^c_\alpha$ do not annihilate in general
this state! The Fock state is related to the free case without
interaction, and only perturbative methods may use it. We want
to calculate bound states, so a strong interaction contradicts
the Fock state.

Because we are interested especially in energy values, we use
$\vert a>$ as an energy eigenstate. $\vert 0>$ is a special case
in which we are able to use Wightman's theorem. In the energy
representation we are able to separate the time dependence.

The (anti) symmetrization in the field operators corresponds to
the Hamilton formalism, because this would be the correct
onetime limiting case if we would deal with a single class of
fields only. To collect the whole set of transition matrix
elements into one expression, we define functional sources $j,b$
and the corresponding contractions $\partial,\partial^b$  with
the relations \index{Functional Fock space}
\be
\{\partial_{I_1},j_{I_2}\}_+&=&\delta_{I_1I_2}\nn
{}[\partial^b_{K_1},b_{K_2}]_-&=&\delta_{K_1K_2}\nn
\partial_{I}\vert 0>_f=\partial^b_{K}\vert0>_f&=&0\nn
{}_f<0\vert j_{I}={\,}_f<0\vert b_{K}&=&0.
\ee
(Not displayed (anti) commutators are zero.)
The functional Fock space construction $\vert 0>_f$ is
independent of the physical vacuum. Now we are able to build the
formal construct of a functional state
\cite{Lur}\index{Generating functional}
\be
\vert{\bf A}(a,j,b)>&:=&\sum_{i=0}^{\infty}
\frac{i^n}{n!m!}\tau^t_{nm}(a)
j^t_{I_1}\ldots j^t_{I_n}b^t_{K_1}\ldots b^t_{K_m}\vert 0>_f\nn
\tau^t_{nm}(a)&=&<0\vert{\cal A}(\Psi^t_{I_1}\ldots\Psi^t_{I_n})
                        {\cal S}(B^t_{K_1}\ldots B^t_{K_m})\vert a>
\ee
where we omit in further formulas the index $t$ for the sake of
brevity. The transition matrix elements are recovered by
`projecting' from the left with the contractions $\partial_{I}$
and $\partial^b_{K}$ and setting the sources zero, which is
equivalent to the action of ${}_f<0\vert$ from the left
\index{Functional `projection'}
\be
\tau^t_{nm}(a)&=&\frac{1}{i^n}{}_f<0\vert\partial^b_{K_1}\ldots
\partial^b_{K_m}\partial_{I_n}\ldots\partial_{I_1}\vert{\bf A}(a,j,b)>.
\ee
Hence one functional state comprises the whole hierarchy, which
can be explicitly recovered by such a projection technique.

\subsection{Dynamics in the Functional Space}

The next step is to implement the dynamics into the functional
formalism. First of all, we have the Hamiltonian corresponding
to the equations (\ref{dyn}) as \index{Hamiltonian of spinor
QED}
\be\label{HPsiB}
H(\Psi ,B)&=&\frac{1}{2}A_{I_1I_3}D_{I_3I_2}\Psi_{I_1}\Psi_{I_2}
            +\frac{1}{2}A_{I_1I_3}W^K_{I_3I_2}B_K\Psi_{I_1}\Psi_{I_2}\nn
          &&+\frac{1}{4}A_{I_1I_5}U_{I_5}^{I_2I_3I_4}
             \Psi_{I_1}\Psi_{I_2}\Psi_{I_3}\Psi_{I_4}\nn
          &&+\frac{1}{2}C_{K_1K_3}L_{K_3K_2}B_{K_1}B_{K_2}.
\ee
There is no term $J_K^{I_1I_2}$ in the Hamiltonian. This term
was eliminated with help of the commutator (\ref{comm-rel}--ii).
Taking the time derivative and using the dynamics, we are left
with \index{Consistency condition, QED}
\be\label{cons}
C_{K_1K}W^K_{I_1I_2}\Psi_{I_2}&=&2A_{I_1I}J^{II_2}_K\Psi_{I_2}.
\ee
This requirement, which is in fact an additional assumption,
that this equation holds for every $\Psi_{I}$ allows to drop the
$\Psi$. Of course $\Psi$ is {\it not} an invertible operator!
This sort of consistency relation may be motivated by the
physical assumption, that there is no asymmetry in the
boson--fermion and fermion--boson interaction (actio equals
reactio, Newton). Furthermore, one has two possibilities of
multiplications of fields operators to antisymmetric groups of
them, that is from right or left. The requirement, that the
result should not depend on that choice also leads to the
above relation \cite{StuFauPfi}.

As a consequence of the introduction of generating functional
states we have to transform (\ref{HPsiB}) into this space
\be
H[\Psi,B]&\Longrightarrow&H[j,b,\partial,\partial^b].
\ee
If we choose $\vert a>$ as energy eigenstate, this results in the
functional Schr\"o\-ding\-er equation \index{Functional
Schr\"odinger equantion}
\be
i\partial_0\vert{\bf A}(a,j,b)>&=&E_{0a}\vert{\bf A}(a,j,b)>=
  H[j,b,\partial,\partial^b]\vert{\bf A}(a,j,b)>.
\ee
In general the matrix elements are transition matrix elements,
and thus not directly related to a statistical interpretation.
The connection to such an interpretation, i.e. the map of the
(positive definite) Hilbert space into functional space, was
investigated in \cite{Stat}.

We are exclusively interested in one time functionals, and may
derive the time derivative by application of the
Baker--Campbell--Hausdorff formula \cite{Mag,Pfi-dip}
\be
i\partial_0 e^{A(t)}&=&\sum_{k=0}^\infty \frac{1}{(k+1)!}
[H,A]_k\nn
{}[H,A]_k&:=&[H,[H,A]_{k-1}];\quad k\geq 1\nn
{}[H,A]_0&:=&i\partial_0 A=[H,A].
\ee
One has to calculate the iterated commutators. Because of the
grade lowering character of the commutators and the
commutativity of bosons and fermions, this process is finite.
The highest term is of order four. The result is an equation
involving functional sources and field operators, i.e. a mixture
of old and new variables
\be
H[\Psi,B]&\Longrightarrow&H[j,b,\Psi,B].
\ee
To eliminate the field operators from this expression, one has
to use the Baker--Campbell--Hausdorff formula once more, with
respect to the functional derivatives (contractions). The final
result is
\be\label{funct-1}
E_{0a}\vert{\bf A}(a,j,b)>&=&\big\{D_{I_1I_2}j_{I_1}\partial_{I_2}
+W^K_{I_1I_2}j_{I_1}\partial_{I_2}\partial^b_K
+L_{K_1K_2}b_{K_1}\partial^b_{K_2}\nn
&&+U_{I_1}^{I_2I_3I_4}j_{I_1}(\partial_{I_2}\partial_{I_3}\partial_{I_4}
  -\frac{1}{4}A_{I_3I_3^\prime}A_{I_2I_2^\prime}j_{I_2^\prime}j_{I_3^\prime}
  j_{I_4})\nn
&&+J_K^{I_1I_2}b_K(\partial_{I_1}\partial_{I_2}+\frac{1}{4}A_{I_1I_1^\prime}
  A_{I_2I_2^\prime}j_{I_1^\prime}j_{I_2^\prime})\big\}\vert{\bf A}
  (a,j,b)>,
\ee
where the $A_{I_nI_m}$ terms stem from the field quantization of
the fermions. Boson quantization terms would again occur, if
the $J_K^{I_1I_2}$ would be eliminated in the above mentioned way.

Summarizing we have obtained the Hamilton formalism transformed
into functional space. The use of the Hamilton formalism is
imperative in order to preserve scalar products and
normalization, which is needed for the derivation of a composite
particle dynamics.

\subsection{Definition of the Composite}

Now we want to define composites to build up positronium states.
Because of the elimination of the superfluous degrees of
freedom, the remaining bosonic coordinates $B_K$ are transverse.
They describe ``real'' photons, in the language of QFT. The
``virtual'' photons are contained in the scalar part, which
was eliminated in favor of the (inter fermionic) Coulomb
interaction. Thus we may restrict ourselves to the fermionic
part of the functional equation. This is easily done by setting
the bosonic sources zero. The remaining equation is a nonlinear
spinor field equation with a nonlocal self coupling.

Before we can proceed, we have to eliminate uncorrelated parts
from the matrix elements. Because we are interested in two
particle bound states, it is sufficient to remove the
uncorrelated parts only up to this stage. This can be achieved
by the functional normal ordering \cite{StuBor}. We thus define
a new functional state \index{Fuctional normal ordering}
\be
\vert{\bf A}(a,j,b)>&=:&e^{-\frac{1}{2}jFj}\vert{\bf F}(a,j,b)>,
\ee
where $F$ is the propagator of the theory and the exponential is
an abbreviation for the formal series.

Inserting this equation into (\ref{funct-1}), commuting the
exponential factor to the left and multiplying with the
inverse from the left, we get the normal ordered equation. This normal
ordering is nonperturbative and {\it not} related to Fock space.
But in the case of Fock space one obtains the same result as in
the usual normal ordering \cite{Wick}, which justifies the name.
We have
\be
\mbox{d}_{I}&:=&\partial_I-F_{II^\prime}j_{I^\prime}\nn
H[j,b,\partial,\partial^b]e^{-\frac{1}{2}jFj}&=&
e^{-\frac{1}{2}jFj}H[j,b,\mbox{d},\partial^b]\nn
&=&e^{-\frac{1}{2}jFj}\hat{H}[j,b,\partial,\partial^b]
\ee
and then (see below (\ref{funct-2}))
\be
E_{0a}\vert{\bf
F}(a,j,b)>&=&\hat{H}[j,b,\partial,\partial^b]\vert{\bf F}(a,j,b)>.
\ee
So far we have normal transformed the full functional equation.
Now we close the bosonic channels in order to define bound
states. This is necessary, because of the meta stable nature of
the positronium states, which would not allow to perform such a
calculation. \index{Definition of composites, heuristic}
\medskip

\noindent
{\bf Def.:} {\it The composite is a solution of the diagonal
part of the functional equation for $b_K,\partial^b_K\equiv 0$.
The diagonal part consists of an equal number
of fermionic sources and contractions without the
renormalization terms.}
\medskip

By this definition the diagonal equation results as
\be\label{diag}
E_{0a}\vert{\bf F}(a,j,b)>&=&\hat{H}^d\vert{\bf F}(a,j,b)>\nn
\hat{H}^d&\approx&j\partial, jj\partial\partial, \ldots
\ee
In configuration space this equation decomposes into a set of
decoupled equations for various lepton numbers. For the
positronium states, we restrict ourselves to the two particle
sector. The formal solution is denoted by
\be
C_k^{I_1I_2}&:=&<0\vert {\cal A}(\Psi_{I_1}\Psi_{I_2})\vert a>,
\ee
which solves (\ref{diag}). $k$ describes the quantum numbers of
the bound state with respect to the state $\vert a>$. The
equations for the diagonal part are very similar to
Bethe--Salpeter like equations.

\subsection{Map onto the Effective Theory}

The ``weak mapping'' \index{Weak mapping} is now a rewriting of
the normal ordered functional equation. This can be achieved by
exact mapping theorems \cite{Map-theorems}. The resulting
equations describe a positronium gas coupled to the electrons,
positrons and photons. All exchange forces and quantum effects
are accounted for.

Here we want to use a short cut method, which is much more
simpler. We neglect thereby several exchange forces, but in the
low energy regime and for small positronium concentrations these
terms become small corrections. A further simplification will be
the restriction to the even part of the functional equation.
This can be done, because the even and odd parts decouple. We
are thus left with the positronium gas and the photon field, as
well as with their coupling and quantum effects.

In the next step we define a set of dual functions to the
positronium solutions $C_k^{I_1I_2}$ by \index{Dual functions of
composites}
\be\label{dual}
R^k_{I_1I_2}C_{k^\prime}^{I_1I_2}&=&\delta^k_{k^\prime}\nn
R^k_{I_1I_2}C_k^{I_3I_4}&=&\delta^{\{I_3I_4\}}_{\{I_1I_2\}_{as}}.
\ee
This duality relation is only possible, if the solutions
$C^{I_1I_2}_k$ span the whole space of two--particles solutions.
We have thus to include scattering states also, which occur if
the rest frame energy increases over a certain threshold. If one
would like to omit these scattering states by the introduction
of an energy cut--off, one would have to project onto a
subspace. However, this would also restrict the possible
physical situations after the mapping.

By means of (\ref{dual}) we introduce new (independent) sources
for the positronium states \index{Positronium sources}
\be\label{source}
c_k&:=&\frac{1}{2}C_k^{I_1I_2}j_{I_1}j_{I_2}.
\ee
The main problem to solve in the above mentioned weak mapping
theorems is the correct definition of the sources $c_k$ as
independent quantities and the correct definition of the corresponding
contraction. This can be achieved exactly. For a rewriting we may
also use, as a short cut, the functional chain rule, given by
\index{Functional chain rule}
\be\label{chain}
\partial_{I_1}&=&\frac{\partial c_k}{\partial j_{I_1}}\partial^c_k=
C_k^{I_1I_2}j_{I_2}\partial^c_k.
\ee
With help of the duals (\ref{dual}), the source definition
(\ref{source}) and the chain rule (\ref{chain}), we can rewrite
the normal ordered functional Hamiltonian
\be
\hat{H}[j,b,\partial,\partial^b]&\cong&
     H[c,b,\partial^c,\partial^b].
\ee
We have to rewrite also the functional state $\vert {\bf
F}(a,j,b)>_{\mbox{even}}$ to $\vert{\bf F}(a,c,b)>$. The resulting
equation is \index{Functional equation, positronium gas}
\be
E_{0a}\vert{\bf F}(a,c,b)>&=&\Big\{
 R^{k_1}_{I_1I_2} D_{I_1I_3} C_{k_2}^{I_3I_2} c_{k_1} \partial^c_{k_2}
-R^k_{I_1I_2} Z_{I_1I_3} F_{I_3I_2} c_k
+L_{K_1K_2} b_{K_1} \partial^b_{K_2} \nn
&&\mbox{\hspace{-2.75cm}}
+W^K_{I_1I_2} \big[ R^{k_1}_{I_1I_3} C_{k_2}^{I_2I_3}
 c_{k_1} \partial^c_{k_2} \partial^b_K
-R^k_{I_1I_3} F_{I_2I_3} c_k \partial^b_K \big] \nn
&&\mbox{\hspace{-2.75cm}}
-U_J^{I_1I_2I_3} \Big[
 4R^{k_1}_{\{JI_4} R^{k_2}_{I_5I_6\}_{as}} C_{k_1^\prime}^{I_1I_4}
 C_{k_2^\prime}^{I_2I_5} C_{k_3^\prime}^{I_3I_6}
 c_{k_1} c_{k_2} \partial^c_{k_1^\prime} \partial^c_{k_2^\prime}
 \partial^c_{k_3^\prime}\nn
&&\mbox{\hspace{-2.75cm}}
-3R^k_{JI_4} C^{I_1I_3}_{k_1^\prime} C^{I_2I_4}_{k_2^\prime}
 c_k \partial^c_{k_1^\prime} \partial^c_{k_2^\prime}
+12R^{k_1}_{\{JI_4} R^{k_2}_{I_5I_6\}_{as}} F_{I_3I_4}
 C_{k_1^\prime}^{I_1I_5} C_{k_2^\prime}^{I_2I_6}
 c_{k_1} c_{k_2} \partial^c_{k_1^\prime} \partial^c_{k_2^\prime}\nn
&&\mbox{\hspace{-2.75cm}}
-3R^{k_1}_{JI_4} F_{I_3I_4} C_{k_2}^{I_1I_2} c_{k_1} \partial^c_{k_2}
-4(F_{I_3I_4} F_{I_2I_5}+\frac{1}{4}A_{I_3I_4} A_{I_2I_5})
 F_{I_1I_6} R^{k_1}_{\{JI_4} R^{k_2}_{I_5I_6\}_{as}} c_{k_1}
 c_{k_2}\nn
&&\mbox{\hspace{-2.75cm}}
+(3F_{I_3I_4} F_{I_2I_5}+\frac{1}{4}A_{I_3I_4} A_{I_2I_5})
 R^{k_1}_{\{JI_4} R^{k_2}_{I_5I_6\}_{as}} C_{k^\prime}^{I_1I_6}
 c_{k_1} c_{k_2} \partial^c_{k^\prime} \Big]\nn
&&\mbox{\hspace{-2.75cm}}
+J_K^{I_1I_2} b_K \big[
 (F_{I_1I_3} F_{I_2I_4}+\frac{1}{4}A_{I_1I_3} A_{I_2I_4})
 R^k_{I_3I_4} c_k
-2R^{k_1}_{I_3I_4} F_{I_1I_3} C_{k_2}^{I_2I_4} c_{k_1} \partial^c_{k_2}\nn
&&\mbox{\hspace{-2.75cm}}
-R^k_{I_3I_4} C_{k_1}^{I_2I_3} C_{k_2}^{I_1I_4} c_k \partial^c_{k_1}
 \partial^c_{k_2}+C_k^{I_2I_1} \partial^c_k \big] \Big\}
 \vert{\bf F}(a,c,b)>.
\ee
If one likes to look at the lowest order of this infinite
hierarchy, one can interpret the functional sources and
contractions as destruction and creation operators and thus as
`images' of the original field operators. Thus a Feynman like
interpretation of the terms is possible. Nevertheless, the
correct evaluation has to use projections and thereby the full
matrix hierarchy. For a brief account on the physical
interpretation of the result we use the former. For example, the
$c\partial^c$ terms constitute a positronium current. A decay
process has $c\partial^b$ and a creation process $b\partial^c$
as functional sources. The coupling of a positronium current to
the boson field is given by $c\partial^c\partial^b$ or
$bc\partial^c$. In spite of the stable bound states, which were
derived from the self--interacting fermion fields, the mapping
reproduces all the necessary effects as decay and creation. As
was mentioned above, the solutions $C^{I_1I_2}_k$ and their
duals $R^k_{I_1I_2}$ contain not only bound states, but also
scattering states. Hence there is the possibility of
dissociation into free electrons and positrons. Because of the
restriction to the even part of the hierarchy, we have no
dynamical theory for the free electrons and positrons, which
nevertheless could be obtained by the complete mapping. Higher
order terms correspond to quantum effects. The concrete
elaboration of the formal result is currently under
investigation \cite{Schar}. One should be able to calculate
numerical data, as the decay constant etc.

\section{Generating Functionals and Clifford Algebras}

Because of the very compact and easy to handle notation of the
functional space, one would like to better understand this
construction. In the above method, only formal aspects of the
functional states were used. The manipulations as normal
ordering and so on were motivated by QFT arguments. This section
is devoted to give a first account on a pure geometric view of
field quantization involved in such states. It will be shown,
that it is not very difficult to give the functional states a
geometric meaning, but the heuristic definition of composites
seems to have no such foundation.

\subsection{New Transition into the Functional Space}

If we want to establish a connection between Clifford algebras
and conventional QFT's, we have to search for a common
structure. This structure is provided by the generating
functionals. Despite their introduction as an auxiliary or mere
book--keeping entity, we want to give them an algebraic and
geometric meaning by the identification with Clifford
algebras. Therefore we look at the generating functionals in a
new way. Only if the geometric interpretation fits somehow into
the QFT framework, this sort of consideration is a reasonable
one. For brevity we restrict ourselves to fermions, but bosons
can be treated along the same lines too. We define the femionic
functional as
\be\label{distri}
\vert{\bf A}(a,j)>&:=&\sum_{n=0}^\infty\frac{i^n}{n!}\tau_n(a)
j_{I_1}\ldots j_{I_n}\vert 0>_f\nn
\tau^t_n(a)=\tau_n(a)&=&i^{-n}{}_f<0\vert\partial_{I_n}\ldots
\partial_{I_1}\vert{\bf A}(a,j)>\nn
&:=&<0\vert{\cal A}(\Psi_{I_1}\ldots\Psi_{I_n})\vert a>.
\ee
The anticommuting sources $j_I$ build up an infinite dimensional
Grassmann algebra. The commutation relations are
\be
\{j_{I_1},j_{I_2}\}_+&=&0.
\ee
Thus we may define in a formal manner the linear space spanned
by this sources
\be
{\cal V}&:=&<j_{I_1},\ldots j_{I_n},\ldots >.
\ee
This set is not countable because of the occurrence of
the space time coordinates in the index $I$. Nevertheless, we
proceed with the formal construction. We expect this sort of
algebra not to exist, because of the renormalization problem of
QFT's as for example mass and charge renormalization in QED.
However by this analogy we want to establish Clifford algebraic
methods in the field theoretic context. This methods should then
give hints how to construct finite field theories, which would
then also exist as a Clifford algebra.

The coefficients of the generating elements $j_I$ of the
Grassmann algebra are complex valued generalized functions, eg.
distributions ($\tau$--functions (\ref{distri})). Thus we have
to build an infinite dimensional Grassmann algebra over the
complex valued generalized functions as \index{Functionals in
Grassmann algebra}
\be
\Lambda{\cal V}&:=&\oplus_n\Lambda^n{\cal V}=D\oplus{\cal V}\oplus
\Lambda^2{\cal V}\oplus\ldots
\ee
with $D$ the space of c-valued generalized functions. Because
even the product of such functions is in general not defined,
one should restrict the domain or has to smear out the fields.
Because of our formal treatment, we do not discuss this topics.
Obviously we then have
\be
\vert{\bf A}(a,j)>&\in&\Lambda{\cal V}
\ee
and the functional states are formal elements of the above
Grassmann algebra.

If we want to translate the `projection' of components out of the
functional states, we have to define contractions. They are
nothing but covectors to the generating elements $j_I$, which
may be called vectors. A vector is thus an element of Grassmann
grade one and a covector is a one form. It is natural to choose
\be
(\partial_{I_1}j_{I_2})&:=&\delta_{I_1I_2}.
\ee
With this definition, we are immediately able to follow
Chevalley \cite{Chevalley} and build the Clifford algebra as a
deformation of the Grassmann algebra. Thus we define
\be
e_I&:=&\partial_I+j_I
\ee
as Chevalley did for vectors over {\bf R} or {\bf C} by
$e_x:=\partial_x +L_x$. Here $\partial_x$ is the usual
derivation and $L_x$ is the exterior (Grassmann) product. In our
case we get
\be
\{e_{I_1},e_{I_2}\}_+&=&2\delta_{I_1I_2}\nn
e^2_I&=&1.
\ee
Of course this equation is singular, and has to be interpreted
as an integral kernel. We have thus constructed the Clifford algebra
\be
CL_D({\cal V},\delta)&=&CL_D(<j_{I_1},\ldots,j_{I_n},\ldots>,\delta)
\ee
as algebra over the c-valued generalized functions, see for an
example \cite{GelShi}. As a consequence, the functional states
are also elements of this Clifford algebra \index{Functionals in
Clifford algebra}
\be
\vert{\bf A}(a,j)>&\in&CL_D({\cal V},\delta).
\ee
The next task is to find a representation of the field operators.
Because of the operator character one has to expect them in the
space of endomorphisms of the Grassmann algebra. Thus we have
\be
\Psi_I&\in&End(\Lambda{\cal V}).
\ee
It is by no means clear, that $\Psi_I$ should be contained in
the Clifford algebra, which is only a subalgebra of
$End(\Lambda{\cal V})$. But this can be concluded from the
canonical commutation relations. We have to notice, that the
commutator is (Grassmann) grade lowering by two. Thus from the
canonical commutation relations
\be
\{\Psi_{I_1},\Psi_{I_2}\}_+&=&A_{I_1I_2}
\ee
we see that $\Psi_I$ is in the Clifford subalgebra of
$End(\Lambda{\cal V})$. With the Ansatz \index{Functional image
of field operators}
\be\label{ansatz}
\Psi_{I_1}&\approx&\frac{1}{i}\partial_{I_1}+
\frac{i}{2}A_{I_1I_2}j_{I_2}
\ee
one can derive the commutator quite easily. In fact this is the
formula one got with the Baker--Campbell--Hausdorff formula of
(left) derivation of the functional state with respect to $j_I$
($\Psi_{I_1}$ is understood here as multiplication operator
inside the $\tau$--functions, from the left)
\be
\partial_{I_1}\vert{\bf A}(a,j)>&=&\frac{\partial}{\partial j_{I_1}}
e^{i\tau j}\vert 0>_f
=(i\Psi_{I_1}+\frac{1}{2}A_{I_1I_2}j_{I_2})\vert{\bf A}(a,j)>.
\ee
The same consideration is possible for boson fields also. In the
same manner we may define the right multiplication of $\Psi_I$
to the $\tau_n(a)$ functions, which is given by
\be
<0\vert{\cal A}
(\Psi_{I_1}\ldots\Psi_{I_n})\Psi_I\vert a>&=&
<0\vert
\frac{1}{i^n}\partial_{I_n}\ldots\partial_{I_1}
(\frac{1}{i}\partial_I-\frac{i}{2}A_{II^\prime}j_{I^\prime})
\vert{\bf A}(a,j)>.
\ee
Therewith we can formulate the identities \index{Functional
identities}
\be
<0\vert\Psi_{J_1}\ldots\Psi_{J_s}B_{L_1}\ldots B_{L_r}
{\cal A}(\Psi_{I_1}\ldots\Psi_{I_n})
{\cal S}(B_{K_1}\ldots B_{K_m})\vert a>&=&\nn
&&\hspace{-8cm}
\frac{1}{i^n}{}_f<0\vert\partial^b_{K_1}\ldots\partial^b_{K_m}
\partial_{I_n}\ldots\partial_{I_1}\times\nn
&&\hspace{-8cm}\times
(\frac{1}{i}\partial^b_{L_1}+\frac{1}{2}C_{L_1L_1^\prime}b_{L_1^\prime})
\ldots
(\frac{1}{i}\partial^b_{L_r}+\frac{1}{2}C_{L_rL_r^\prime}b_{L_r^\prime})
\times\nn
&&\hspace{-8cm}\times
(\frac{1}{i}\partial_{J_s}+\frac{i}{2}A_{J_sJ_s^\prime}j_{J_s^\prime})
\ldots
(\frac{1}{i}\partial_{J_1}+\frac{i}{2}A_{J_1J_1^\prime}j_{J_1^\prime})
\vert{\bf A}(a,j,b)>
\ee
and
\be
<0\vert\
{\cal A}(\Psi_{I_1}\ldots\Psi_{I_n})
{\cal S}(B_{K_1}\ldots B_{K_m})\vert a>
\Psi_{J_1}\ldots\Psi_{J_s}B_{L_1}\ldots B_{L_r}
&=&\nn
&&\hspace{-8cm}
\frac{1}{i^n}{}_f<0\vert\partial^b_{K_1}\ldots\partial^b_{K_m}
\partial_{I_n}\ldots\partial_{I_1}\times\nn
&&\hspace{-8cm}\times
(\frac{1}{i}\partial^b_{L_1}-\frac{1}{2}C_{L_1L_1^\prime}b_{L_1^\prime})
\ldots
(\frac{1}{i}\partial^b_{L_r}-\frac{1}{2}C_{L_rL_r^\prime}b_{L_r^\prime})
\times\nn
&&\hspace{-8cm}\times
(\frac{1}{i}\partial_{J_s}-\frac{i}{2}A_{J_sJ_s^\prime}j_{J_s^\prime})
\ldots
(\frac{1}{i}\partial_{J_1}-\frac{i}{2}A_{J_1J_1^\prime}j_{J_1^\prime})
\vert{\bf A}(a,j,b)>
\ee
with
\be
A_{I_1I_2}&:=&\{\Psi_{I_1},\Psi_{I_2}\}_+\nn
C_{K_1K_2}&:=&[B_{K_1},B_{K_2}]_-.
\ee
The proof can be done by iteration of single applications of
$\Psi_I$ and $B_K$ to the left and right. Such formulas were
derived in \cite{Pfi-dip,StuFauPfi,StuBor} \rule{1ex}{1ex}.

By means of these identities on can get in a very easy way the
functional Schr\"o\-ding\-er equation. Because of
\be
i\partial_0\tau_{nm}(a)&=&i\partial_0<0\vert
{\cal A}(\Psi_{I_1}\ldots\Psi_{I_n})
{\cal S}(B_{K_1}\ldots B_{K_m}\vert a>\nn
&=&<0\vert\Big[
H[\Psi,B],
{\cal A}(\Psi_{I_1}\ldots\Psi_{I_n}){\cal S}(B_{K_1}\ldots B_{K_m})
\Big]\vert a>\nn
&=&<0\vert H[\Psi,B]
{\cal A}(\Psi_{I_1}\ldots\Psi_{I_n}){\cal S}(B_{K_1}\ldots B_{K_m})
\vert a>\nn
&&-
<0\vert
{\cal A}(\Psi_{I_1}\ldots\Psi_{I_n}){\cal S}(B_{K_1}\ldots B_{K_m})
H[\Psi,B]\vert a>
\ee
we have the fields inside the Hamiltonian one time left and one
time right of the (anti) symmetrized blocks. Thus we get the transition
\be
H[\Psi,B]&\Longrightarrow& H[j,b,\partial,\partial^b]
\ee
by subtracting the two different substitutions of $H[\Psi,B]$
with $\Psi\rightarrow\partial\pm Aj$, $B\rightarrow\partial^b
\pm Cb$ as
\be
H[j,b,\partial,\partial^b]&=&H[\Psi,B]\big\vert_{
\Psi\approx\partial+Aj \atop B\approx\partial+Cb}
-H[\Psi,B]\big\vert_{
\Psi\approx\partial-Aj \atop B\approx\partial-Cb}.
\ee
The resulting functional equation is equivalent to (\ref{funct-1}).
Besides the enormous shortened calculation, we have established
that the commutators of field quantized theories are nothing
but the metric of the desired Clifford algebra in the functional
space. One yields therefore
\be
\vert{\bf A}(a,j,b)>&\in&CL_D({\cal V}\oplus{\cal W},A\oplus C)
\ee
were ${\cal V}$ is the fermion (vector) space and ${\cal W}$ is
the boson (vector) space spanned by $j$ and $b$. $A$ and $C$ are
the corresponding metrics.

In a previous paper we investigated the vertex normal ordering,
which has to be performed in a QFT with nonlinear self
interaction \cite{Fau-ver}. There we found the connection
between this reordering and the choice of a specific Clifford
algebra out of a class of isomorphic ones. Even our composite
calculation forced us to introduce a normal ordering. Thus we
have to look for a generalization of the above transition. This
is necessary, because if we choose the normal ordered QFT to be
finite, then the not normal ordered one has singular
interaction terms and vice versa. To avoid such ill defined
steps in the calculation, we have to find a direct transition to
the normal ordered equation.

So far, we had only symmetric $A_{I_1I_2}$ in the
Ansatz (\ref{ansatz}) of the fermion operators. We allow now
arbitrary bilinearforms $B$, which we can split $B$ into a
symmetric part $A$ and an antisymmetric part $F$. Since only $A$
and $F$ occur in functional formulas a confusion with the
bosonfields also denoted $B$ is impossible. We can then
introduce left and right multiplications as \index{Functional
image of normal ordered field operators}
\be\label{PsiAF}
\hat{\Psi}_{I_1}^{L/R}&\approx&\frac{1}{i}\partial_{I_1}\pm
\frac{i}{2}A_{I_1I_2}j_{I_2}+iF_{I_1I_2}j_{I_2}.
\ee
The asymmetric choice of factors $1/2$ is due to physical
considerations about the normal ordering, which was introduced
above by $d_I:=\partial_I-F_{II^\prime}j_{I^\prime}$. Of course
this choice also leads to the correct pre--factors in the normal
ordered functional equation (\ref{funct-2}). An even better
choice would have changed the commutator to $2A$. The $i$
factors are relics of the definition of the fermionic
functionals as $exp(i\tau j)$ and could be doped. Of course we
choose (\ref{PsiAF}) for the reason of comparison with
previously obtained results.

Much more important is the occurrence of the {\it same} sign $(+)$
in front of the $F_{I_1I_2}$. That is, we have in the
antisymmetric part a strong asymmetry in the conjugation
structure. This sort of asymmetry was studied to a considerable
length in \cite{Fau-cli}. Hence we remark, that the right and
left action is distinguished by the {\it commutator}, while the
propagator acts both times in the same way. This corresponds to a
non grade invariant involution, which has the propagator as
scalar part (in the sense of Grassmann grade). Of course this
stems from the use of the naive Grassmann wedge built up from
antisymmetric Grassmann sources of definite Grassmann grade,
i.e. the $j_I$. Because the injection $\Lambda{\cal
V}\rightarrow CL({\cal V},A)$ is by no means unique, the normal
ordering is nothing but the transformation to the correct, that
is the physical reasonable, wedge. For a detailed account on
that see \cite{Fau-ver}. Thus the {\it propagator} fixes the
involution and therefore the grading structure. This choice is
thus nothing but the choice of the appropriate physical vacuum,
which is here represented as the part with (Grassmann) grade
zero, belonging to the `physical' choice from above.

We have the commutators for the bosons \index{Functional
commutators}
\be\label{comm-funct}
{}[\partial^b_{K_1}-\frac{1}{2}C_{K_1K_1^\prime}b_{K_1^\prime},
\partial^b_{K_2}-\frac{1}{2}C_{K_2K_2^\prime}b_{K_2^\prime}]_-&=&
C_{K_1K_2}\nn
{}[\partial^b_{K_1}+\frac{1}{2}C_{K_1K_1^\prime}b_{K_1^\prime},
\partial^b_{K_2}+\frac{1}{2}C_{K_2K_2^\prime}b_{K_2^\prime}]_-&=&
-C_{K_1K_2}\nn
{}[\partial^b_{K_1}\pm\frac{1}{2}C_{K_1K_1^\prime}b_{K_1^\prime},
\partial^b_{K_2}\mp\frac{1}{2}C_{K_2K_2^\prime}b_{K_2^\prime}]_-&=&0,
\ee
while we have for the normal ordered fermions
\be
\{\frac{1}{i}\partial_{I_1}\pm\frac{i}{2}A_{I_1I_1^\prime}j_{I_1^\prime}
+iF_{I_1I_1^\prime}j_{I_1^\prime},
\frac{1}{i}\partial_{I_2}\pm\frac{i}{2}A_{I_2I_2^\prime}j_{I_2^\prime}
+iF_{I_2I_2^\prime}j_{I_2^\prime}\}_+&=&\nn
&&\hspace{-10cm}
\pm\frac{1}{2}A_{I_2I_2^\prime}\{\partial_{I_1},j_{I_2^\prime}\}_+
+F_{I_2I_2^\prime}\{\partial_{I_1},j_{I_2^\prime}\}_+
\pm\frac{1}{2}A_{I_1I_1^\prime}\{\partial_{I_1^\prime},j_{I_2}\}_+
+F_{I_1I_1^\prime}\{\partial_{I_1^\prime},j_{I_2}\}_+\nn
&=&
\pm A_{I_1I_2}\nn
\{\frac{1}{i}\partial_{I_1}\pm\frac{i}{2}A_{I_1I_1^\prime}j_{I_1^\prime}
+iF_{I_1I_1^\prime}j_{I_1^\prime},
\frac{1}{i}\partial_{I_2}\mp\frac{i}{2}A_{I_2I_2^\prime}j_{I_2^\prime}
+iF_{I_2I_2^\prime}j_{I_2^\prime}\}_+&=&0.
\ee
If one had chosen $\pm F$ as for the commutator, one would not
have reached the goal of left and right multiplication. This is,
because of the mixed commutator, which would have not been zero.
But left and right multiplications have to be commutative
operations in any associative algebra.

We are now able to calculate in a direct manner the (fermion)
normal ordered functional Hamiltonian from $H[\Psi,B]$.
Therefore we have to build
\be
\hspace{-0.75cm}
\hat{H}[j,b,\partial,\partial^b]&=&H[\Psi,B]\vert_{
\Psi_I\approx\frac{1}{i}\partial_I+\frac{i}{2}A_{II^\prime}j_{I^\prime}
+iF_{II^\prime}j_{I^\prime} \atop
 B_K\approx\partial^b_K+\frac{1}{2}C_{KK^\prime}b_{K^\prime}}
-
H[\Psi,B]\vert_{
\Psi_I\approx\frac{1}{i}\partial_I-\frac{i}{2}A_{II^\prime}j_{I^\prime}
+iF_{II^\prime}j_{I^\prime} \atop
 B_K\approx\partial^b_K-\frac{1}{2}C_{KK^\prime}b_{K^\prime}}.
\ee
We have according to (\ref{HPsiB}) four terms $T_i$. They will be
now separately considered
\be
T_1&=&\frac{1}{2}A_{I_1I_3}D_{I_3I_2}\Psi_{I_1}\Psi_{I_2}\nn
&\approx&
\frac{1}{2}A_{I_1I_3}D_{I_3I_2}\Big[
(\frac{1}{i}\partial_{I_1}+\frac{i}{2}A_{I_1I_1^\prime}j_{I_1^\prime}
+iF_{I_1I_1^\prime}j_{I_1^\prime})
(\frac{1}{i}\partial_{I_2}+\frac{i}{2}A_{I_2I_2^\prime}j_{I_2^\prime}
+iF_{I_2I_2^\prime}j_{I_2^\prime})\nn
&&-
(\frac{1}{i}\partial_{I_1}-\frac{i}{2}A_{I_1I_1^\prime}j_{I_1^\prime}
+iF_{I_1I_1^\prime}j_{I_1^\prime})
(\frac{1}{i}\partial_{I_2}-\frac{i}{2}A_{I_2I_2^\prime}j_{I_2^\prime}
+iF_{I_2I_2^\prime}j_{I_2^\prime})\Big]\nn
&=&
\frac{1}{2}A_{I_1I_3}D_{I_3I_2}\Big[\big\{
(\frac{1}{i}\partial_{I_1}+\frac{i}{2}A_{I_1I_1^\prime}j_{I_1^\prime})
(\frac{1}{i}\partial_{I_2}+\frac{i}{2}A_{I_2I_2^\prime}j_{I_2^\prime})
\nn
&&-
(\frac{1}{i}\partial_{I_1}-\frac{i}{2}A_{I_1I_1^\prime}j_{I_1^\prime})
(\frac{1}{i}\partial_{I_2}-\frac{i}{2}A_{I_2I_2^\prime}j_{I_2^\prime})
\big\}\nn
&&+\big\{
(\frac{1}{i}\partial_{I_1}+\frac{i}{2}A_{I_1I_1^\prime}j_{I_1^\prime})
iF_{I_2I_2^\prime}j_{I_2^\prime}
+
iF_{I_1I_1^\prime}j_{I_1^\prime}
(\frac{1}{i}\partial_{I_2}+\frac{i}{2}A_{I_2I_2^\prime}j_{I_2^\prime})
\nn
&&-
(\frac{1}{i}\partial_{I_1}-\frac{i}{2}A_{I_1I_1^\prime}j_{I_1^\prime})
iF_{I_2I_2^\prime}j_{I_2^\prime}
-
iF_{I_1I_1^\prime}j_{I_1^\prime}
(\frac{1}{i}\partial_{I_2}-\frac{i}{2}A_{I_2I_2^\prime}j_{I_2^\prime})
\big\}\nn
&&+\big\{
-F_{I_1I_1^\prime}F_{I_2I_2^\prime}j_{I_1^\prime}j_{I_1^\prime}
+F_{I_1I_1^\prime}F_{I_2I_2^\prime}j_{I_1^\prime}j_{I_1^\prime}
\big\}
\Big].
\ee
With help of symmetry arguments on $\partial$ and $j$ as the
antisymmetry of the $AD$ factor in $I_1I_2$ we have
\be
T_1&=&\frac{1}{2}A_{I_1I_3}D_{I_3I_2}\Big\{
4\frac{i}{2}\frac{1}{i}A_{I_1I_1^\prime}j_{I_1^\prime}\partial_{I_2}
+4i\frac{i}{2}A_{I_1I_1^\prime}F_{I_2I_2^\prime}
j_{I_1^\prime}j_{I_2^\prime}+0\Big\}\nn
&=&D_{I_1I_2}j_{I_1}\partial_{I_2}
-D_{I_1I_3}F_{I_3I_2}j_{I_1}j_{I_2},
\ee
which yields because of the nature of the propagator and the
renormalization discussed in \cite{StuFauPfi}
\be
T_1&=&D_{I_1I_2}j_{I_1}\partial_{I_2}
-Z_{I_1I_3}F_{I_3I_2}j_{I_1}j_{I_2}\nn
Z_{I_1I_3}&:=&-\delta m\gamma^0_{\alpha_1\alpha_3}\delta_{\Lambda_1
\Lambda_3}\delta({\bf r}_1-{\bf r}_3).
\ee
The term $T_2$ is given as
\be
T_2&=&\frac{1}{2}A_{I_1I_3}W^K_{I_3I_2}B_K\Psi_{I_1}\Psi_{I_2}\nn
&\approx&
\frac{1}{2}A_{I_1I_3}W^K_{I_3I_2}\Big[
(\partial^b_K+\frac{1}{2}C_{KK^\prime}b_{K^\prime})(\ldots
)(\ldots )\nn
&&-(\partial^b_K-\frac{1}{2}C_{KK^\prime}b_{K^\prime})(\ldots
)(\ldots )\Big]\nn
&=&
\frac{1}{2}A_{I_1I_3}W^K_{I_3I_2}\Big[\partial^b_K\big\{
(\ldots)(\ldots )-(\ldots)(\ldots )\big\}\nn
&&+\frac{1}{2}C_{KK^\prime}b_{K^\prime}\big\{
(\ldots)(\ldots )+(\ldots)(\ldots )\big\}
\Big].
\ee
The first term is just the same as before, while the second
yields the terms
\be
\mbox{2. brace}\quad
\{\}&=&-2\partial_{I_1}\partial_{I_2}-2\frac{1}{4}A_{I_1I_1^\prime}
A_{I_2I_2^\prime}j_{I_1^\prime}j_{I_2^\prime}\nn
&&+4F_{I_1I_1^\prime}j_{I_1^\prime}\partial_{I_2}
-2F_{I_1I_1^\prime}
F_{I_2I_2^\prime}j_{I_1^\prime}j_{I_2^\prime}.
\ee
Further more we may use (\ref{cons}) to obtain
\be
\frac{1}{4}A_{I_1I_3}W^K_{I_3I_2}C_{KK^\prime}b_{K^\prime}
&=&
-\frac{1}{4}A_{I_1I_3}C_{K^\prime K}b_{K^\prime}W^K_{I_3I_2}\\
&=&
-\frac{1}{2}A_{I_1I_3}A_{I_3I_4}J^{I_4I_2}_{K^\prime}b_{K^\prime}\nn
&=&
-\frac{1}{2}J^{I_1I_2}_K b_K
\ee
and obtain as result
\be
T_2&=&W^K_{I_!I_2}\big[
j_{I_1}\partial_{I_2}-F_{I_2I_2^\prime}j_{I_1}j_{I_2^\prime}\big]
\partial^b_K\nn
&&+J_k^{I_1I_2}b_K\big[
\partial_{I_1}\partial_{I_2}
-2F_{I_1I_1^\prime}j_{I_1^\prime}\partial_{I_2}\nn
&&
+(F_{I_1I_1^\prime}F_{I_2I_2^\prime}+\frac{1}{4}
 A_{I_1I_1^\prime}A_{I_2I_2^\prime})j_{I_1^\prime}j_{I_2^\prime}
 \big].
\ee
The third term is lengthy and will be omitted, as the scheme is
clear now. The last term runs along the lines of $T_1$ and has
not changed at all from (\ref{funct-1}). Our result is then
\be\label{funct-2}
E_{0a}\vert{\bf F}(a,j,b)>&=&\Big\{
 D_{I_1I_2}j_{I_1}\partial_{I_2}-Z_{I_1I_3}F_{I_3I_2}j_{I_1}j_{I_2}\nn
&&
+W^K_{I_!I_2}\big[
j_{I_1}\partial_{I_2}-F_{I_2I_2^\prime}j_{I_1}j_{I_2^\prime}\big]
\partial^b_K\nn
&&
+J_k^{I_1I_2}b_K\big[
 \partial_{I_1}\partial_{I_2}
 -2F_{I_1I_1^\prime}j_{I_1^\prime}\partial_{I_2}\nn
&&+(F_{I_1I_1^\prime}F_{I_2I_2^\prime}+\frac{1}{4}
    A_{I_1I_1^\prime}A_{I_2I_2^\prime})j_{I_1^\prime}j_{I_2^\prime}
    \big]\nn
&&
+U_{J}^{I_1I_2I_3}j_{J}
\big[
\partial_{I_1}\partial_{I_2}\partial_{I_3}
-3F_{I_3I_4}j_{I_4}\partial_{I_2}\partial_{I_1}\nn
&&
+(3F_{I_3I_4}F_{I_2I_5}+\frac{1}{4}
   A_{I_3I_4}A_{I_2I_5})j_{I_4}j_{I_5}\partial_{I_1}\nn
&&
-(F_{I_3I_4}F_{I_2I_5}F_{I_1I_6}+\frac{1}{4}
  A_{I_3I_4}A_{I_2I_5}A_{I_1I_6})j_{I_4}j_{I_5}j_{I_6}\big]\nn
&&
+L_{K_1K_2}b_{K_1}\partial^b_{K_2}
\Big\}\vert{\bf F}(a,j,b)>.
\ee
which coincides with (8.59) in \cite{StuBor} and (63) in
\cite{StuFauPfi}. So we have derived the normal ordered
functional equation in one single step by Clifford algebraic
considerations. No normal ordering problems, as a vertex normal
ordering, was necessary. No singular intermediate equation has
occurred.

\subsection{Criticism of the Heuristic Composite Definition}

Our next step should be to define the composite. But if we want
to translate simply the concept of taking the diagonal part, we
do not succeed. Of course it is possible to extract the diagonal
equations quite easily in the Clifford algebraic picture too,
but there are no geometric arguments to do so. Thus we have to
reject this sort of considerations in the algebraic framework.
This is a new argument against Bethe--Salpeter like equations.

The next problem is, that even in the QFT formalism one can not
be satisfied with such ``hard core'' solutions. There should be
a polarization cloud consisting of infinite many
particle--anti--particle pairs. A translation of such a
polarization cloud into the geometric picture would result in a
highly nonlinear equation.
The same problem occurs if one likes to use selfconsistency
arguments. This stems from the fact that a part of the
renormalized particle is built up from ``vacuum'' polarization.
The Clifford algebraic picture unveils thus this wrong point of
view. The physical ``vacuum'' was simply wrong chosen by the use
of the naive (Grassmann) grading. The injection of the Grassmann
grading into the Clifford algebra has carefully to be
accounted for and the ``vacuum'' is the scalar part of the
nonsymmetric involution \cite{Fau-ver,Fau-cli}.

New $C^\ast$--algebraic investigations \cite{Ker} give a hint,
that a $q$--deformed theory {\it for the composites} may solve
this problem. In such cases one has to choose $q$ as a root of
unity. The representations of this algebras are reducible and
offer some freedom to build up new states. A Clifford algebraic
formulation of this will be given elsewhere.
%%%
%%% H. Stumpf
%%%
One can also remark that in general composite particle dynamics
can not be described by a (usual, i.e. canonically quantized)
quantum field theory at all. The use of $q$--deformations
restricts the situation too, which may be useful in some
situations but which does not cover the whole range of composite
particle physics as will be shown elsewhere.
%%%
So, as a result there are {\it no} current methods to define QFT
bound states in an algebraic framework, but the clear Clifford
algebraic formulation gives some hints and provides the tools
how to build them.

\section{Summary}

In the first section we gave an outline of the well developed method
of weak mapping. This method provides a most useful and
consistent approach to QFT via generating functionals. It also
offers the best available mechanism to deal with composites
including their quantum properties and dynamics. As an example
we treated spinor QED in Coulomb gauge to demonstrate the method.
The considerations are fully contained in the QFT picture
of matrix elements, operators, field quantization, states and so
on. The occurring problems might be handled by heuristic methods
as for example the renormalization procedure.

In the second section we tried to give a new interpretation of
field quantization. This was motivated by Clifford algebraic
methods. As a surprising fact we had to distinguish between
several isomorphic Clifford algebras parameterized by the
propagator of the theory. We can summarize this in the following
conjecture:\index{Conjecture, geometricity of field
quantization}\medskip

{\it Field quantization is nothing but the introduction of
geometry into the functional space. The {\bf commutator} chooses the
desired Clifford algebra, the {\bf propagator} fixes the specific
representant of the isomorphic algebras.}\medskip

We have thus learned to use geometric arguments instead of
a heuristic `quantization' of classical theories. The geometric
reasoning did provide us a direct way to the normal ordered
functional equation without passing through ill defined
intermediate steps.

The heuristic composite definition was rejected by this sort of
geometric arguments. But the method itself gives a hint how to
proceed with composite definitions. This will be done in detail
in another work.

%\appendix
%\section{}

\section*{Acknowledgements}
One of the authors (B.F.) would like to thank the organizers of
the ICTE '95 for the invitation and their hospitality, a
participation would have not be possible without financial support.
Special thanks will be given to Prof. Z. Oziewicz for very
illuminating discussions concerning covectors.
\vskip 0em plus 1em minus 1 em


\begin{thebibliography}{99}
%
\bibitem{Operator-pro} T. Holstein and H. Primakoff {\it Field
dependence of the intrinsic domain magnetization of a
ferromagnet}, Phys. Rev. 58 1940 1098--1113\par
F.J. Dyson {\it Thermodynamic behavior of
an ideal ferromagnet}, Phys. Rev. 102 1956 1230--1244\par
M. Gell-Mann and K.A. Bruekner {\it Correlation energy of an
electron gas at high density}, Phys. Rev 106 1957 364--368\par
T. Usui {\it Excitations in a high density electron gas I.},
Prog. Theor. Phys. 23 1960 787--798
%
\bibitem{BetSal} E. Salpeter and H. Bethe, {\it A Relativistic
Equation for Bound--State Problems}, Phys. Rev. 84 1951
1232--1242\par
H.A. Salpeter and E.E. Salpeter, {\it Quantum mechanics of one--
and two--electron atoms}, Plenum Publ. Corp. New York 1977\par
see also the algebraic work of W. Krolikowski, {\it Clifford
algebras and algebraic structures of fermions}, 171--181 in
proc. of the second Max Born Symposium 1992 ed. Z. Oziewicz and
B. Jancewiec, Kluver 1993
%
\bibitem{BSFlaw} K. Hideji and W. Masami {\it On the
Bethe--Salpeter Equation and the Exited States of the Hydrogen
Atom}, Prog. Theor. Phys. 17(1) 1957 63--79
%
\bibitem{StuBor} H. Stumpf and Th. Borne, {\it Composite Particle Dynamics
in Quantum Field Theory}, Vieweg 1994
%
\bibitem{QED-neu} H. Stumpf and W. Pfister {\it Algebraic
treatment of quantum electrodynamics in temporal gauge},
submitted to Annals of Physics (New York)
%
\bibitem{Dir-ham} P.A.M. Dirac,  {\it Lectures on quantum field
theory}, Belfer Graduate School of Science, Yeshiva Univ. New
York 1966
%
\bibitem{rel-inv} P. Besting and D. Sch\"utte, {\it Relativistic
invariance of Coulomb--gauge Yang--Mills theory}, Phys.
Rev D 42 1990 594--601\par
S.K. Kim, W. Namgung, K.S. Soh and J.H. Yee {\it Equivalence
between the covariant, Weyl and Coulomb gauges in the functional
Schr\"odinger picture}, Phys. Rev.
D 43 1991 2046--2049
%
\bibitem{StuFauPfi} H. Stumpf, B. Fauser and W. Pfister, {\it
Composite Particle Theory in Quantum Electrodynamics}, Z.
Naturforsch 48a 1993 765--776
%
\bibitem{Lee-QED} T.D. Lee and C.N. Yang {\it Theory of Charged
Vector Mesons Interacting with the Electromagnetic Field}, Phys. Rev.
128(2) 1962 885--898\par
N.H Christ and T.D. Lee {\it Operator ordering
and Feynman rules in gauge theories}, Phys. Rev. D 22(4) 1980
939--958
%
\bibitem{Wig} A.S. Wightman, {\it Quantum field theory in terms
of vacuum expectation values}, Phys. Rev. 101 1956 860--866
%
\bibitem{Lur} D. Luri\'e, {\it Particles and fields}, John Wiley
\& Sons, New York 1968\par
see also \cite{StuBor} and references therein. Functionals were
used extensively in the Heisenberg school, for example by
Freese, Symanzik etc. The use of functional methods can be traced
back to Volterra.
%
\bibitem{Stat} H. Stumpf, {\it Normalization and probability
densities for wave functions of quantum field theoretic energy
equations}, Z. Naturforsch. 44a 1989 262--268
%
\bibitem{Mag} W. Magnus, {\it On the exponential solution of
differential equations for a linear operator}, Comm. on Pure and
Appl. Math. Vol. VII 1954 649--637
%
\bibitem{Pfi-dip} W. Pfister, {\it Der einzeitige Formalismus in
der funktionalen Quantentheorie}, diploma thesis,
University of T\"ubingen 1987
%
\bibitem{Wick}
F.J. Dyson, {\it The Radiation Theories of Tomonaga, Schwinger,
and Feynman}, Phys. Rev. 75 1949 486--502\par
F.J. Dyson, {\it The S--Matrix in electrodynamics}, Phys. Rev.
75 1949 1736--1755\par
G.C. Wick, {\it The evaluation of the collision
matrix}, Phys. Rev. 80 1950 268--272\par
F.J. Dyson, {\it Divergence of Perturbation Theory in Quantum
Electrodynamics}, Phys. Rev. 85 1952 631--632
%
\bibitem{Map-theorems} R. Kerschner and H. Stumpf, {\it
Derivation of composite particle dynamics by weak mapping of
quantum fields}, Ann. Phys. (Leipzig) 48 1991 56--68\par
W. Pfister and H. Stumpf {\it Exchange Forces of Composite
Particles in Quantum Field Theory}, Z. Naturforsch. 46a 1991
389--400\par
see also \cite{StuBor}
%
\bibitem{Schar} J. Schaarschmidt, Thesis, University
of T\"ubingen, in preparation
%
\bibitem{Chevalley} C. Chevalley, {\it The algebraic theory of
spinors}, Columbia University Press 1954
\par
Z. Oziewicz {\it From Grassmann to Clifford}, p. 245--256 in
Proceedings of ``Clifford Algebras and their Application in
Mathematical Physics'' Canterbury 1985, Kluver 1986
%
\bibitem{GelShi} I.M. Gelfand and G.E. Schilow, {\it
Verallgemeinerte Funktionen (Distributionen) I}, VEB Deutscher
Verlag der Wissenschaften 1960, Chap. 3\par
A.S. Wightman and L. Garding, {\it Fields as operator valued
Distributions in relativistic quantum field theories}, Ark. Fys.
23(13) 1964
%
\bibitem{Fau-ver} B. Fauser, {\it Vertex Normalordering as a
Consequence of Nonsymmetric Bilinearforms in Clifford Algebras},
hep-th/9504055, accepted for publ. in J. Math. Phys.
%
\bibitem{Fau-cli} B. Fauser, {\it Clifford Algebraic Remark on
the Mandelbrot Set of Two--Component Number Systems},
hep-th/9507133, submitted to Adv. in Appl. Clifford Alg.
%
\bibitem{Ker} R. Kerschner, {\it On quantum field theories with
finitely many degrees of freedom}, hep-th/9505063
%
\end{thebibliography}
\end{document}